\documentclass{article}
\usepackage{spconf,amsmath,graphicx,hyperref}

\usepackage{booktabs}

\title{Contrastive timbre representations for musical instrument and synthesizer retrieval}
\twoauthors
  {Gwendal Le Vaillant\sthanks{This work was also supported by the IRISIB research institute.}}
	{Haute-École Bruxelles-Brabant\\
	ISIB\\
	Rue Royale 150, 1000 Brussels, Belgium}
  {Yannick Molle\sthanks{Funding: Wallonia Region - WIN4SO program.}}
	{University of Mons\\
	ISIA Lab, Pulgn\\
	Bd Dolez 31, 7000 Mons, Belgium}

\begin{document}
\maketitle
\begin{abstract}

Efficiently retrieving specific instrument timbres from audio mixtures remains a challenge in digital music production. This paper introduces a contrastive learning framework for musical instrument retrieval, 
enabling direct querying of instrument databases using a single model for both single- and multi-instrument sounds. 
We propose techniques to generate realistic positive/negative pairs of sounds for virtual musical instruments, such as samplers and synthesizers, addressing limitations in common audio data augmentation methods. 

The first experiment focuses on instrument retrieval from a dataset of 3,884 instruments, using single-instrument audio as input. Contrastive approaches are competitive with previous works based on classification pre-training.
The second experiment considers multi-instrument retrieval with a mixture of instruments as audio input.
In this case, the proposed contrastive framework outperforms related works, achieving 81.7\% top-1 and 95.7\% top-5 accuracies for three-instrument mixtures.

\end{abstract}

\section{Introduction}
\label{sec:intro}

Modern digital music production, driven by digital audio workstations and virtual instruments (samplers and synthesizers), offers vast timbral possibilities.
However, efficiently selecting the right instrument remains challenging. 
Listening to many different instruments is impractical and time-consuming, and textual descriptions of instruments, such as their name and category, often fail to capture timbre nuances.

\begin{figure}[t]
    \centering
    \includegraphics[width=1.0\linewidth]{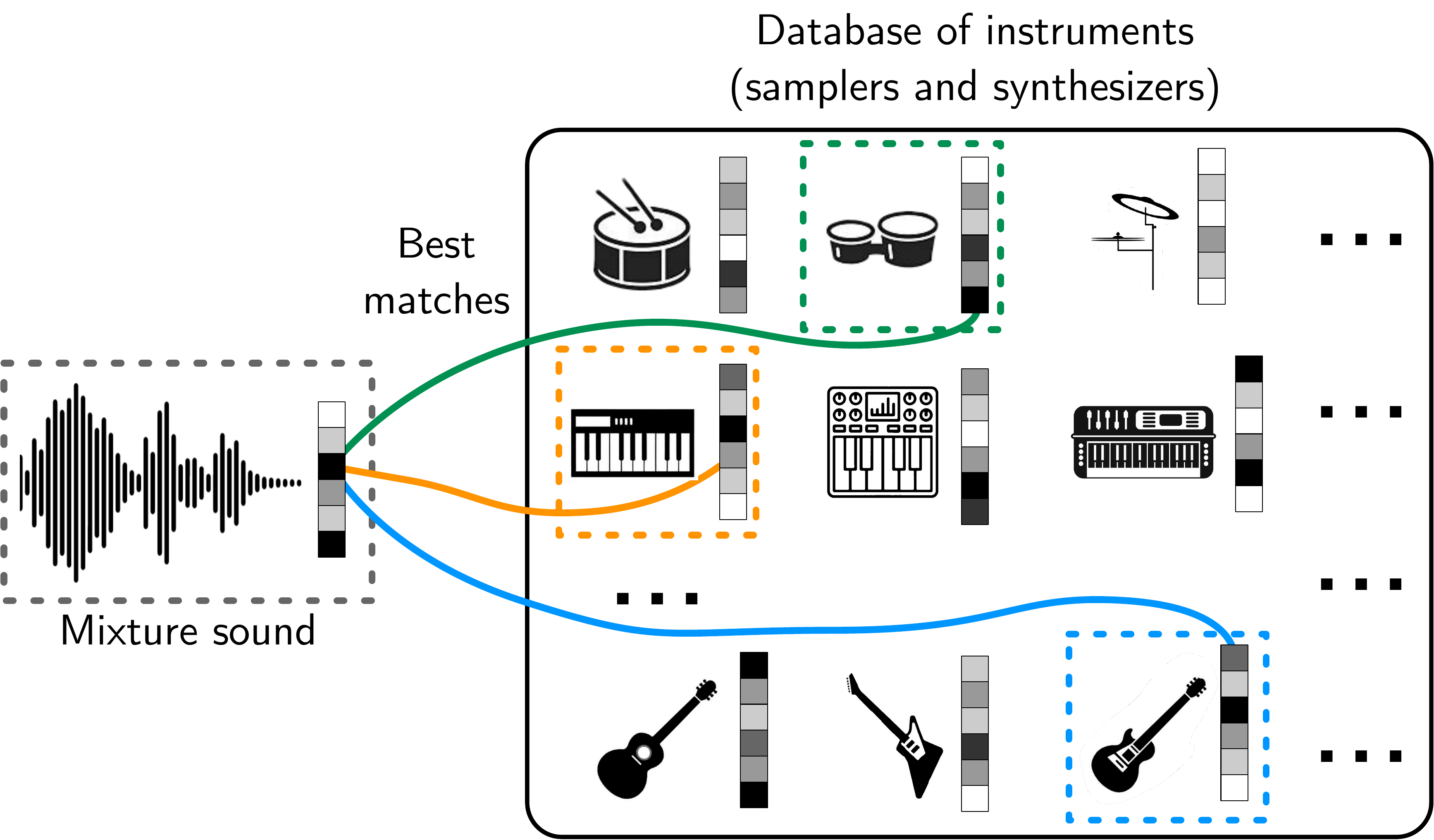}
    \caption{Musical instrument retrieval from a mixture, based on embeddings cosine similarity. All embeddings (single- and multi-instrument sounds) are computed by a unique model introduced in this paper.}
    \label{fig:embedding_retrieval}
\end{figure}

This paper focuses on musical instrument retrieval, which consists in finding a specific instrument's timbre within a mixture of several instruments~\cite{instru_retrieval_mixture_ICASSP23}. 
Timbre is defined as the set of auditory attributes that allow listeners to identify sound sources~\cite{timbre_5_dims_jasa2013, timbre_MIR_vs_music_psycho_jnmr2016}, but it is hard to quantify~\cite{timbre_5_dims_jasa2013, timbre_hard_to_understand_krumhansl89}.
This instrument retrieval task is more challenging than family classification with broad categories (e.g., \textit{bass}, \textit{flute}, \textit{strings}).
In particular, our work introduces a model able to compute embeddings for both single- and multi-instrument sounds, which allows querying instruments in a database (Fig.~\ref{fig:embedding_retrieval}).
This unique model is trained to discriminate between instruments using a contrastive objective.
A substantial number of recent works have used contrastive objectives for speech processing tasks, such as speaker identification and verification~\cite{contrastive_speaker_embedding_TASLP24, SSL_speaker_verif_segments_ICASSP24}.
However, contrastive approaches for mixes of musical instruments and synthesizers have not been formally studied yet.

The main contributions of this paper are threefold:
1) dedicated techniques to build positive and negative pairs of sounds for the contrastive objectives;
2) a comparison of related works for instrument retrieval from single-source and multi-source audio, with interpretable results provided as \mbox{top-1} and \mbox{top-5} accuracies for a large merged dataset of 3884 instruments;  
and 3) a contrastive model for musical instrument retrieval, reporting $81.7$\% \mbox{top-1} and $95.7$\% \mbox{top-5} accuracies for mixtures of three instruments, outperforming related works by a large margin.

\section{Related works}
\label{sec:related_works}

\subsection{Contrastive audio representations}

Contrastive audio representation learning aims to encode meaningful embeddings from raw unlabelled audio by maximizing agreement between positive pairs, which typically are augmented views of the same dataset sample, while minimizing similarity for negative pairs (usually different samples).
Data augmentation can rely on crops and translations in time, small frequency shifts of input spectrograms, and added noise~\cite{semantic_triplet_loss_ICASSP18}, while large datasets such as AudioSet~\cite{audioset_ICASSP2017} allow generating millions of pairs without requiring explicit labels.
Models can be pre-trained using dedicated losses, e.g., triplet~\cite{facenet_triplet_loss_CVPR15} or Information Noise-Contrastive Estimation (InfoNCE, \cite{CPC_InfoNCE_arXiv19}) losses.
Downstream tasks include 
speech or sound classification~\cite{semantic_triplet_loss_ICASSP18, COLA_ICASSP21},
or query-by-example (QbE) \cite{semantic_triplet_loss_ICASSP18}.

For multi-source musical audio, \cite{multi_source_contrastive_SCM23} combines source separation~\cite{open_unmix_JOSS19}, also called stem splitting, 
with contrastive learning techniques~\cite{COLA_ICASSP21}.
In particular, \cite{multi_source_contrastive_SCM23} optimizes the similarity of augmented pairs extracted from an automatically extracted stem: the vocal, bass, drums, or \textit{other} track.
It is competitive with related works for downstream tasks such as instrument or genre classification, but has not been formally evaluated for instrument retrieval in a piece of music.

Contrastive language-audio pretraining~\cite{CLAP_LAION_ICASSP23} has also demonstrated impressive results on various downstream tasks, although this indeed requires labeled or annotated samples.
In the context of instrument retrieval, in particular for synthesizers, instrument names are not always available, or can be inconsistent and noisy. They can even be random names, not describing the instrument's timbre at all. 
Hence, language-audio models have not been further considered in this paper.

\subsection{Instrument retrieval}

Using previous works, a straightforward way to retrieve the closest instrument from a database of instruments is based on embeddings computed by pretrained models. Given a mixture of instruments, multiple stems (typically four) can be approximately extracted. Then, an audio embedding can be computed for one of the stems.
Finally, the retrieved instrument is the one whose audio embedding best matches the stem embedding, based on a cosine or L2 distance for instance.
Nonetheless, stem splitting remains a limitation for this method.
First, it typically restricts separation to broad categories, lumping diverse sounds like all synthesizers into an undifferentiated \textit{other} stem~\cite{open_unmix_JOSS19, demucs_ISMIR21}.
Second, it introduces artifacts such as distortions and bleeding, which impair the retrieval.

A recent work~\cite{instru_retrieval_mixture_ICASSP23} proposes to use two models in parallel: one is trained to compute embeddings of single-instrument sounds, whereas the other computes multiple embeddings for each mixture sound.
The single-source model is first pretrained for NSynth~\cite{nsynth_wavenet_icml2017} instrument classification, then a permutation-invariant cosine distance loss is used to ensure that the multi-source model computes embeddings similar to the target single-source embeddings.
The multi-source model can finally be used to query a database of instruments' embeddings.
Nonetheless, \cite{instru_retrieval_mixture_ICASSP23} does not consider contrastive approaches.

\section{Contrastive timbre embeddings}
\label{sec:single_source}

\subsection{Motivation}

Similar to applications in computer vision~\cite{SimCLR_ICML20}, previous works~\cite{semantic_triplet_loss_ICASSP18, COLA_ICASSP21, multi_source_contrastive_SCM23} on contrastive audio representations rely on data augmentation in the time (cropping) or spectral (frequency shift) domains, performed on a unique audio excerpt.
Noise or reverberation can also be used~\cite{SSL_speaker_verif_segments_ICASSP24}. 
However, such methods do not seem the most appropriate for building representations of the timbre of an instrument.
For instance, a sound made of a unique note would be split into two sounds considered as a positive pair. The \textit{attack} of the sounds, one of the most salient timbre descriptors~\cite{timbre_MIR_vs_music_psycho_jnmr2016, timbre_hard_to_understand_krumhansl89, timbretoolbox_jasa11}, would be separated from the \textit{sustain} and  \textit{release} parts of the note. Each of those two sounds would miss important aspects of timbre.

In practice, some spectro-temporal descriptors of timbre can co-vary with pitch, loudness or duration~\cite{timbre_MIR_vs_music_psycho_jnmr2016}. Hence, simple spectrogram augmentations such as gains or frequency shifts are not adapted. 
Noisiness is also a timbre attribute~\cite{timbre_5_dims_jasa2013, timbretoolbox_jasa11} and should not be used for data augmentation.
In this paper, we use the virtual instruments themselves to generate appropriate and realistic pairs of sounds for contrastive training (Fig.~\ref{fig:dataset}).
This section focuses on timbre embeddings for single-source sounds.
Section~\ref{sec:mixed_sources} handles the case of contrastive embeddings for mixtures.

\subsection{Dataset and sampling strategies}\label{ssec:single-source-dataset-sampling}

We build a large dataset based on Nsynth~\cite{nsynth_wavenet_icml2017} (1000 instruments) and the Surge~\cite{surge_synthesizer} 
synthesizer (2884 \textit{patches}, considered as instruments).
NSynth provides around 300,000 audio files (330h of audio). Instruments are classified into eleven families corresponding to MIDI instrument families such as \textit{bass}, \textit{percussion}, \textit{strings}, \textit{brass}, \textit{synth lead}, etc. 

\begin{figure}[t]
    \centering
    \includegraphics[width=1.0\linewidth]{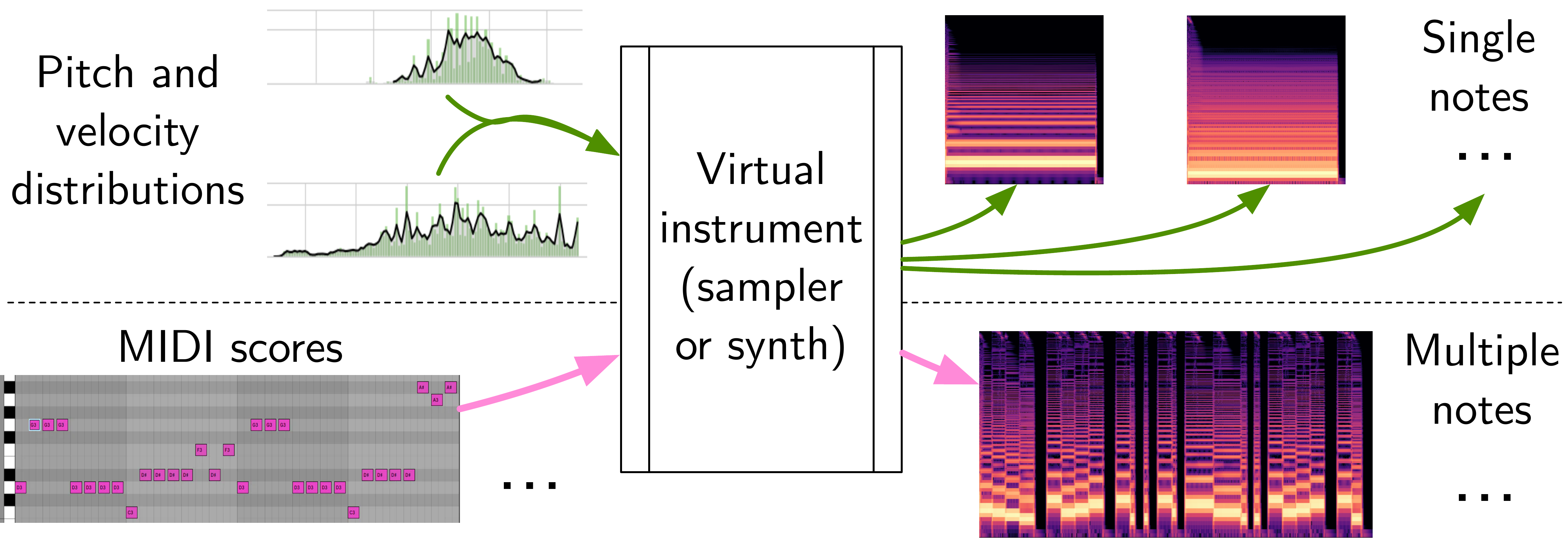}
    \caption{Dataset generation for a virtual instrument, using family-specific MIDI distributions and scores extracted from Slakh~\cite{slakh_dataset_IEEEWASPAA19}. All sounds in this figure are generated by the same instrument, thus any pair is considered positive.}
    \label{fig:dataset}
\end{figure}

Surge instruments are classified into 25 categories. Most of them could be associated to an NSynth instrument family, or to the \textit{synth pad} family. 
A total of twelve MIDI families is then considered for our dataset.
To generate audio files from Surge instruments, the Slakh dataset~\cite{slakh_dataset_IEEEWASPAA19} is useful, thanks to its MIDI scores separated by instrument family.
These scores were used to generate 10s-long audio files (Fig.~\ref{fig:dataset}).
The empirical marginal distributions of MIDI pitch and velocity for each family were also estimated, 
and used to generate single-note audios files (Fig.~\ref{fig:dataset}), similar to NSynth's notes.
In contrast to NSynth which samples pitches and velocities uniformly for each instrument, leading to degenerated timbres for very high or low pitches, the use of family-specific MIDI data from Slakh allows us to generate consistent audio samples for each instrument.

Some Surge patches also use various audio effects (reverberation, delay, phaser, distortion, ...). 
In this study, we consider that removing these effects is an alteration of timbre, hence it is a data augmentation technique.
Some 2123 augmented instruments, obtained by disabling effects on original training patches, are appended to the training dataset.
This consistently improved the performance of all models presented in the following experiments.
The final generated Surge dataset contains 4422 training, 285 validation and 284 test instruments, with a total of approximately 200,000 files (300h of audio).

\subsection{Model, training and evaluation}\label{ssec:single_source_model_training_eval}

The contrastive training process relies on an Audio Spectrogram Transformer (AST,~\cite{AST_interspeech21}) for computing embeddings from Mel-spectrograms. The AST is pretrained on AudioSet~\cite{audioset_ICASSP2017} using a classification head. 
For training with minibatch size $N$, we first sample $N/2$ different instruments, then draw a pair of positive audio files for each.
Given a positive pair, all sounds from other instruments are considered as negatives.
We train with a minibatch size $N=24$.

After training, AST embeddings can be used for QbE evaluation on the test set.
To emulate a realistic and difficult task, we first build a database of 3864 candidate embeddings using all training, validation and test instruments (augmented instruments excluded).
For each instrument, a single embedding is computed using the observed median pitch and velocity,
which helps reduce the computational cost of evaluation.
Second, all 13000 test sounds (336 instruments) are encoded using the trained AST, and the cosine distances with all database embeddings are computed.
The best matching instruments can finally be retrieved from these distances.

\subsection{Experiment}\label{ssec:single-source-experiments}

Table~\ref{tab:single_source_results} provides the top-1 and top-5 accuracies of various single-source encoders on the test set, and compares the contrastive approach to previous works.
The top section reports performance of the Timbre Toolbox~\cite{timbretoolbox_jasa11} baseline.
We exclude features related to pitch and loudness, and normalize (zero-mean, unit-variance) the remaining timbre attributes to obtain embedding vectors for QbE. 

The middle section of Table~\ref{tab:single_source_results} 
covers two AST models trained on pretext tasks: instrument family 
classification, and direct classification of the training instrument~\cite{instru_retrieval_mixture_ICASSP23}.
After training, the classification layer is discarded, and only the last hidden activations are used as embeddings for instrument retrieval.
The first classifier is trained on only twelve broad and possibly noisy classes.
The second classifier has 5730 output logits corresponding to all training instruments, and seems to build precise internal  representations which transfer well to the test dataset for the QbE task. 
It is the best \textit{single-encoder} model, reused in Sub-sec.~\ref{ssec:mix-experiment} of this paper.

The bottom section of Table~\ref{tab:single_source_results} analyzes two contrastive objectives.
The InfoNCE loss maximizes the similarity of positive pairs, and minimizes that of pairs of sounds from different instruments.
The triplet loss uses each sound once as anchor and once as positive, with all sounds from other instruments considered negatives.
For QbE with single-source queries, contrastive methods slightly underperform relative to previous works~\cite{instru_retrieval_mixture_ICASSP23} but remain competitive.

\begin{table}
    \centering
    \begin{tabular}{r|c|c}
           & Top-1 & Top-5 \\
        \midrule
        Timbre Toolbox \cite{timbretoolbox_jasa11} & $29.7$ & $40.1$ \\
        \midrule
        Instrument family classification & $62.9$ & $75.2$ \\
        Training instrument classification~\cite{instru_retrieval_mixture_ICASSP23} & $\mathbf{83.2}$ & $\mathbf{95.0}$ \\
        \midrule
        Contrastive: triplet loss & $79.0$ &  $91.2$ \\    
        Contrastive: InfoNCE & $80.4$ & $93.1$ \\  
    \end{tabular}
    \caption{QbE test accuracies (\%) of single-source encoders.}
    \label{tab:single_source_results}
\end{table}

\section{Multi-source contrastive training}
\label{sec:mixed_sources}

\subsection{Contrastive objective}

Works mentioned in Sec.~\ref{sec:related_works} are tailored to the task of finding multiple instruments in a mixture.
In this section, we train an AST model in a purely contrastive framework, then apply it to musical instrument retrieval without further fine-tuning.

\begin{figure}[t]
    \centering
    \includegraphics[width=1.0\linewidth]{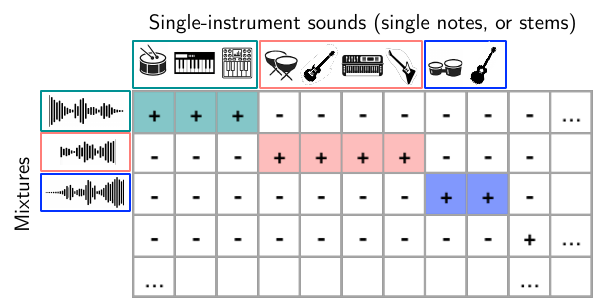}
    \caption{Positive and negative pairs of sounds in a training minibatch. All stems must use different instruments, although families can overlap.}
    \label{fig:mix_pairs}
\end{figure}

The contrastive objective with target similarities is provided in Fig.~\ref{fig:mix_pairs}.
Given mixtures made of different instruments, the model maximizes the similarity between a mixture and its constituents (Fig.~\ref{fig:mix_pairs}), while minimizing the similarity with instruments from other mixtures.
Pairs of mixtures are also considered as negative, which is not represented in Fig.~\ref{fig:mix_pairs} for the sake of conciseness.
This framework handles an arbitrary number of instruments per mixture, but two different training mixtures cannot contain the same instrument.

\subsection{Experiment}\label{ssec:mix-experiment}

\subsubsection{Baselines and dataset}

The first baseline is the Demucs stem splitter~\cite{demucs_ISMIR21, hybrid_transformers_ICASSP23} combined with the best single-source encoder from Table~\ref{tab:single_source_results}.
The comparative experiment from this sub-section is then constrained to mixtures of three instruments: 
one \textit{percussion}, one \textit{bass} and one \textit{synth lead} instrument. The last family falls in the \textit{other} category of Demucs.
Given a Demucs-splitted mixture, the QbE described in Sub-sec.~\ref{ssec:single_source_model_training_eval} can be directly performed separately for each instrument family, using the best single-source encoder from Table~\ref{tab:single_source_results}.

The dataset is then reduced to 1463 instruments: 615 \textit{basses}, 480 \textit{synth leads} and 368 \textit{percussions} (data augmentations excluded). 
For training and evaluation, mixtures are obtained either from stems made of Slakh MIDI scores, or from randomly sampled and concatenated single notes and silences (no chords). We ensure that no mixture component is silent.
The evaluation is performed on 5000 test mixtures generated from held-out test instruments,
and Table~\ref{tab:mix_results} reports the average accuracy of QbE performed separately for each instrument family.
The poor performance of this first baseline suggests that state-of-the-art stem splitting, with its associated artifacts (Sec.~\ref{sec:single_source}), is not yet suitable for precise timbre retrieval.

\begin{table}
    \centering
    \begin{tabular}{r|c|c}
           & Top-1 & Top-5 \\
        \midrule
        Demucs~\cite{demucs_ISMIR21, hybrid_transformers_ICASSP23} + single-encoder & $14.5$ & $25.8$ \\
        Multi-encoder~\cite{instru_retrieval_mixture_ICASSP23} & $17.32$ & $62.6$ \\
        \midrule
        Contrastive: triplet loss & $64.8$ & $85.0$ \\
        Contrastive: full triplet loss & $\mathbf{81.7}$ & $\mathbf{95.7}$ \\
        Contrastive: InfoNCE loss & $70.0$ & $93.3$ \\
    \end{tabular}
    \caption{Average QbE test accuracies (\%) for mixture sounds.}
    \label{tab:mix_results}
\end{table}

Another solution is to train the multi-encoder from~\cite{instru_retrieval_mixture_ICASSP23}, in which we replace the convolutional architecture with a more powerful AST.
We compute a loss without permutations for this constrained experiment, which leads to better performance. 
The multi-encoder generates one embedding for each instrument family, and these embeddings are forced to be close to those computed by the frozen single-encoder.
After training, each output embedding is used for a QbE performed independently for each instrument family.
The $62.6$\% top-5 average accuracy is much better than the baseline, although the top-1 accuracy remains quite low (under $20$\%).

\subsubsection{Contrastive models}

Table~\ref{tab:mix_results} also reports the performance of our contrastive models.
We optimize similarities as indicated previously, and use single-notes as single-instrument sounds (columns of Fig.~\ref{fig:mix_pairs}) rather than the multi-note stems summed to generate the mixtures.
This helps reduce the discrepancy between the training and test setups.
In contrast to previous methods, our approach relies on a unique model to compute all embeddings---for mixtures and for the database of single-instrument sounds.

Different contrastive losses are considered.
The triplet loss uses mixtures as anchors, and the associated single-instrument notes as positives.
Compared to the baselines, it leads to a substantial increase in accuracy.
The \textit{full} triplet loss uses each sound (single-instrument, or mixture) as anchor at some point, which improves the representations even more. This is the best model, with $81.7$\% top-1 and $95.7$\% top-5 accuracies.
Using InfoNCE does not further improve the results, although the performance is consistently better than that of the baselines. Audio examples are available online\footnote{\url{https://gwendal-lv.github.io/CIR}}.

The multi-encoder approach~\cite{instru_retrieval_mixture_ICASSP23} is trained to precisely generate embeddings of training instruments.
Unlike our contrastive approaches, the multi-encoder seems to overfit on embeddings of training timbres, although the training dataset of mixtures is virtually infinite.
The contrastive approach is also more straightforward, and does not require multiple models with different embedding spaces in parallel~\cite{instru_retrieval_mixture_ICASSP23, multi_source_contrastive_SCM23}, or two-step training strategies~\cite{instru_retrieval_mixture_ICASSP23}.

\section{Conclusion}

This paper introduces a contrastive approach for musical instrument retrieval from multi-instrument input mixtures.
A unique model can be used to encode both a mixture and a database of single-instrument sounds with audio effects, in order to query the database by example.
An experiment with mixtures of three instruments demonstrates a large increase in accuracy compared to the previous state of the art.
The contrastive model generates fine timbre representations of both single- and multi-instrument sounds, 
which can be used to retrieve instruments from a user's personal database, or for a specific synthesizer or a given instrument family.
The embeddings could also be used as audio tokens in audio-language frameworks~\cite{CLAP_LAION_ICASSP23, LTU_ICLR24} dedicated to musical instruments.

\vfill\pagebreak

\bibliographystyle{IEEEbib}
\bibliography{refs}

\end{document}